\documentclass[%
 reprint, 
showpacs,
footinbib,
bibnotes,
 amsmath,amssymb,
 aps,
 prl,
]{revtex4-1}

\usepackage[export]{adjustbox}
\usepackage{amsmath,amsfonts,amsthm, amssymb, braket, relsize, exscale, mathtools}
\usepackage{float}
\usepackage{mathrsfs, morefloats}
\usepackage{dsfont}
\usepackage{graphicx}
\usepackage{epstopdf} 
\usepackage[colorinlistoftodos]{todonotes}
\usepackage{setspace}

\usepackage{silence}
\begin{document}

\preprint{APS/123-QED}

\title{High energy photon production in strong colliding laser beams}

\author{M. Yu. Kuchiev} 
 \email[]{  kmy@phys.unsw.edu.au}
\author{J. Ingham}%
 \email{z3374563@unsw.edu.au}
\affiliation{%
University of New South Wales, Sydney
  2052, Australia\\
}%

\date{\today}

\begin{abstract}
The collision of two intense, low-frequency laser beams is considered. The $e^-e^+$ pairs created in this field 
are shown to exhibit recollisions, which take place at high energy accumulated due to the wiggling of fermions.
The resulting $e^-e^+$ annihilation produces high energy photons, or heavy particles. 
The coherent nature of the laser field provides strong enhancement of the probability of these events. Analytical and numerical results are outlined.
\end{abstract}

\maketitle

A strong static electromagnetic field can create electron-positron pairs
as was shown by Sauter and Schwinger
\cite{Sauter,Schwinger1951}. 
Similarly, the phenomenon of the pair creation takes place in slow-varying fields, 
as was demonstrated in
\cite{Narozhnyi1973, Marinov1977, Brezin1970, Yakovlev1964, Keldysh1965}. 
This letter elucidates the role of recollisions of the $e^-e^+$ pairs created in slowly oscillating fields. 
We show that in a standing wave produced by two colliding laser beams the creation and the following wiggling of electrons and positrons in the laser field pave the way for their annihilation with high energy gamma-production, or production of heavy particles. 
The probability of these phenomena is found to be greatly enhanced by the coherent 
nature of the laser field.

The $e^-e^+$ production is exponentially suppressed 
for fields below the Schwinger limit $\mathcal E_c = m^2/e$ ($\hbar=c=1$, if not
stated otherwise).
However, several laser facilities are aiming to produce fields 
within an order of magnitude of $\mathcal E_c $ in the next decade \cite{ELI2011, xfeleu, hiperlaser}.
There were also considered several ways in which the probability  of $e^-e^+$ 
creation may be boosted.
Refs. \cite{Kohlfurst2013, Bell2008} discussed manipulation of the laser pulse shape and polarization. 
Other studies investigated the dynamical Schwinger mechanism \cite{Schutzhold2008}, 
in which a weak field is used to lower the barrier for the strong field tunnelling process. 
Recent work by Di Piazza et. al. \cite{DiPiazza2009} claims that 
by tuning the frequency of the weak field, $e^-e^+$ creation 
may be observable with presently available technology. 
Meuren et. al. \cite{Meuren2015, Meuren2015a} considered collision 
of a single photon with a strong laser pulse when the rate of $e^-e^+$ creation  is enhanced by the high photon energy.

The recollision, or antenna mechanism as originally posed in \cite{Kuchiev2007}, is a three step process. First, an $e^-e^+$ pair is created through the Schwinger mechanism. Second, $e^-$ and $e^+$ oscillate in the laser field, and third, they recollide. These recollisions violate the adiabatic condition, 
and hence are exponentially enhanced compared to conventional adiabatic processes.
In Ref. \cite{Kuchiev2007}, the recollision of $e^-e^+$ pairs created by a heavy nucleus exposed to the laser beam were considered,
while Meuren et. al. \cite{Meuren2015, Meuren2015a} considered recollisions in the field of a strong laser pulse incident on a single photon.
In the problem discussed in the present paper, when only the laser fields operate, the antenna mechanism reveals a new interesting feature, which strongly enhances the collision probability.
The effect originates from the fact that the velocities of the created $e^-$ and $e^+$ are correlated along the direction of the laser electric field.

The low energy analogue of the antenna process is the atomic antenna mechanism proposed in \cite{Kuchiev1987}, and developed quasiclassically by Corkum, and Lewenstein et al. \cite{Corkum1993, Lewenstein1994}, in which high-harmonic radiation, multiple ionization and above threshold ionization take place. This is currently an area of active research \cite{Shwartz2014, Smirnova2009}.

Consider the collision of two laser beams represented by two plane waves propagating head-on along the $x$-direction with the same frequency, intensity, and linear polarization along the $z$-axis. Assume that their phases are tuned to produce the standing wave in which the total electric field is $E=\mathcal E\,\cos\omega t \,\cos kx$.
We presume that the electric field $\mathcal E$, frequency 
$\omega$, and adiabatic parameter $\gamma= m\omega/(e\mathcal E)$ 
are small, $\mathcal E \lesssim \mathcal E_c$, $\omega\ll m$, $\gamma\ll 1$;
the last two inequalities are essential
to the $e^-e^+$ pair acquiring high energies. 

We evaluate the probability $W$ of $e^-e^+$ creation using the Keldysh-type approach \cite{Keldysh1965}, in which 
\begin{align}
W \propto \sum_s \int |A(\mathbf p)|^2 \frac{d^3p}{(2\pi)^3} \ .
\label{W}
\end{align}
Here the integration and summation run over the electron momentum $\mathbf p$ and its spin projection $s$ at the moment
of creation; one finds that the positron momentum equals
$-\mathbf p$ while the projections of the electron and positron
spins along $\mathbf p$ are opposite. 
For the amplitude $A(\mathbf p)$ of the pair creation in the relativistic region $\mathbf p^2\gg m^2$ we find 
\begin{align}
&A(\mathbf p)=A_0 \exp(i S_\text{tun}({\mathbf p}))~,
\label{AiS}
\\
&S_\text{tun}({\mathbf p})=\frac{i\pi}{2F}\,\Big(m^2+{p_\perp^2}+\frac{\gamma^2 p_z^2}{2}\Big)-\frac{p_z^2}{F}~,
\label{S}
\end{align}
where $S_\text{tun}({\mathbf p})$ is the tunneling action
and $F=e\mathcal E$.
Equations (\ref{W})-(\ref{S}) show that typical electron momenta along the electric field greatly exceed perpendicular momenta,
$p_z\propto F^{1/2}/\gamma\gg p_\perp \propto F^{1/2}$.
The created fermions therefore propagate at small angles to the electric field, which complies
with expectations based on physical reasons.
Observe also that (\ref{W})-(\ref{S}) dictate that the pair is created mostly in the vicinity of points where the amplitude of the electric field is maximal, 
$\omega t=\pi n$, $\omega x=\pi m$, for integer $n,m$.  The magnetic field in the standing wave,
$B_y=\mathcal E\,\sin\omega t \,\sin kx$, is small near these points. Correspondingly, on most classical trajectories 
the fermions are exposed only to a small magnetic field and an almost homogeneous electric one.  
Hence below we neglect the magnetic field and treat the electric one as homogeneous.

We aim to calculate the probability of the annihilation of the created pair, $e^-e^+\rightarrow 2\gamma$, 
which occurs over a short time interval compared to the laser period, 
$\omega \delta t_\text{col}\ll 1$. As a result, the annihilation amplitude can be factorized into the matrix element describing the annihilation of two fermions with 
energies $\varepsilon$,
and the amplitude describing the fermion propagation to the point of collision. 
A similar factorization takes place in the known problem of 
the positronium annihilation, see e.g. \cite{Berestetskii1982}.
This analogy allows us to write the following expression for the conversion coefficient, 
which equals the probability of $e^-e^+$ annihilation,
\begin{equation}
\Upsilon\!=\! \frac 2 \pi \int 
|\Phi_{2\varepsilon}(0)|^2\, v \, \sigma(\varepsilon)~ d\varepsilon.
\label{recoll}
\end{equation}
Here $\sigma(\varepsilon)$ is the cross section of the fermion annihilation in the vacuum 
and $v$ is the electron speed at the moment of collision, while $\Phi_{2\varepsilon}(\mathbf r_{12})$ is the wave function of the fermion pair. It appears in (\ref{recoll}) at zero separation between fermions $r_{12}=0$ to allow the annihilation to take place. 

To evaluate $\Phi_{2\varepsilon}(0)$ we first construct  the time dependent wave function of the fermion pair $\Psi(\mathbf r_{12},t)$, and then examine its Fourier transform. 
The function $\Psi(\mathbf r_{12},t)$ needs to take into account that the pair can be created with a given initial electron momentum $\mathbf p$; 
the corresponding amplitude is presented
in (\ref{AiS}). The propagation amplitudes of the fermions are given by
their single particle wave functions $\psi^{ (-) }_{\mathbf{p}}(\mathbf r,t)$
and $\psi^{ (+) }_{-\mathbf{p}}(\mathbf r,t)$, where the supercript $(\mp)$
marks the electron and positron states. 
The pair creation and its following propagation are consequent events. Hence the wave function
reads
\begin{equation}
\Phi({\mathbf r}_{12},t)= \hat P\int 
A(\mathbf p)\psi^{ (-) }_{\mathbf{p}}(\mathbf r_1,t)\psi^{ (+) }_{-\mathbf{p}}(\mathbf r_2,t)\frac{d^3p}{(2\pi)^3},
\label{Apsipsi}
\end{equation}
where the integration takes into account interference of events with 
different momenta, and the operator $\hat P$ chooses the necessary opposing spin states projections of the pair.
We find the single-particle wave functions 
using the adiabatic approximation, which is applicable since $\omega\ll m$. The electron wavefunction reads
\begin{align}
\psi^{ (-) }_{\mathbf{p}}(\mathbf r,t)=
\chi^{ (-) }_{\mathbf{p}(t)} 
\exp\Big(i \big(\,\mathbf{p}(t)\cdot \mathbf r-\int^t\varepsilon(t')dt'\,\big)\Big)~,
\label{e}
\end{align}
where $\chi^{ (-) }_{\mathbf{p}}$  
is the usual spinor that describes the electron propagation with the given momentum $\mathbf p$ in the vacuum.
In the time-dependent electric field this momentum, as well as the energy, become slowly varying functions, 
$\mathbf p(t)=\mathbf p+(\mathbf F/\omega) \sin \omega\,t$, $\varepsilon (t)=(\mathbf p^2(t)+m^2)^{1/2}$. 
The integral in the phase factor, $\int^t\varepsilon(t')dt'$,  is an important and conventional feature of the adiabatic approximation. The positron wavefunction is constructed similar to (\ref{e}), with momentum 
$-\mathbf p(t)$.

Substituting Eqs.(\ref{AiS}), (\ref{e}) into the wavefunction (\ref{Apsipsi}) we find
\begin{align}
&\Psi(\mathbf r_{12},t)= A_0\int
\xi_\mathbf p 
\exp\big(i S(\mathbf p, \mathbf r_{12},t)\big)\,
\frac{d^3p}{(2\pi)^3},
\label{wavefunction}
\\
&S(\mathbf p, \mathbf r_{12},t)=\mathbf{p}(t)\cdot \mathbf r_{12}-2\int^t\varepsilon(t')dt'+
S_\text{tun}(\mathbf p)~,
\label{A=S-E}
\end{align}
where $\xi_\mathbf p =\hat P (\chi^{ (-) }_{\mathbf{p}(t)}\chi^{ (+) }_{-\mathbf{p}(t)})$
is the spin part, 
$S(\mathbf p, \mathbf r_{12},t)$ can be considered a classical action, which incorporates 
the tunneling action that describes the pair creation, 
and the action responsible for its subsequent propagation. 
The constant $A_0$ here is defined by normalization conditions.
Since $m/\omega \gg 1$, the integral in \eqref{wavefunction} may be evaluated by the saddle point method. 
The location of the saddle point is defined by $\nabla_{\mathbf p}S(\mathbf p, \mathbf r_{12},t)=0$, 
in which one recognizes the Hamilton-Jackoby condition for a classical trajectory.
This implies that the saddle-point calculations we fulfill amount to a semiclassical description of the problem.

\begin{figure}[tcb]
{\includegraphics[scale=0.4]{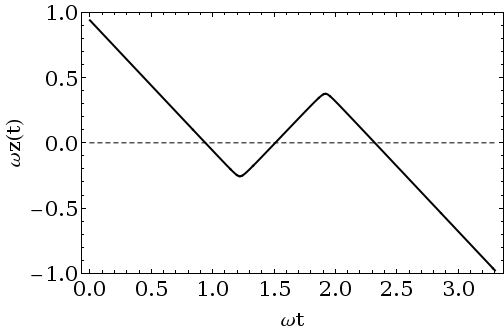}}
\caption{The electron classical trajectory $z(t)$, 
$\gamma=10^{-1.5}$, $p_z=-0.9 F^{1/2}/\gamma$. The nodes of $z(t)$ indicate 
the moments of time where $e^-e^+$ collisions may occur.}
\label{trajectory}
\end{figure}

For the case of $r_{12}=0$ we are interested in, the relevant classical trajectory has 
a simple, appealing form. 
Using (\ref{S}) and (\ref{A=S-E}) we find
that the electron and positron velocities have opposite directions and are strictly aligned along the electric field, so that the electron and positron coordinates 
can be taken as $z(t)$ and 
$-z(t)$ respectively. The saddle point condition dictates
\begin{align}
z(t)=\int_0^t v(t') \,dt'+{\varepsilon(0)}/{F}=0.
\label{trajectory_explicit}
\end{align}
The velocity equals $v(t)=p(t)/(p^2(t)+m^2)^{1/2}$, where
the lower limit of integration in (\ref{trajectory_explicit}) 
indicates that we consider the pair created at $\omega t\approx0$. 
We see that at the moment of creation the fermions are separated by the distance $2z(0)=2{\varepsilon(0)}/{F}\approx 2|p_z|/F$, which is similar to what takes place in the static electric field.
The related term in the middle expression in (\ref{trajectory_explicit})
was derived from (\ref{S}) neglecting corrections $\propto \gamma^2$. 
The last identity in Eq.(\ref{trajectory_explicit}) states that the
necessary condition $\mathbf r_{12}=0$ takes place at the nodes of the function $z(t)$.
Since $p(t)$ oscillates, there may exist several nodes of $z(t)$, i. e. several collision times.
From Eq. (\ref{trajectory_explicit}) we find that these nodes
appear only when $p_z<0$. Three such nodes are present in Fig. \ref{trajectory}, where the trajectory $z(t)$ is shown for $\gamma=10^{-1.5}$ with $p_z=-0.9 F^{1/2}/\gamma$. These nodes are enumerated by $n=0, \ 1, \ 2$ below.

Evaluating the integral in (\ref{wavefunction}) at the saddle point 
(\ref{trajectory_explicit}) we find
\begin{align}
&\Psi(0,t)=\xi_p\, \Phi(0,t)
\label{explicit-wf})
\\
&\Phi(0,t)\approx \sum_{n\ge 0}\frac{\gamma^{1/2}\,F^{3/4}}{2^{7/4}\,D(p)}
\exp \Big(i (S(p,0,t)-\frac {3\pi}4)\Big)
\nonumber
\\
&D(p)=
\begin{cases}
2n (2n+v(0))^{1/2}\,\ln \frac \pi\gamma, &n\ge 1
\\
(1+Y_1(\tfrac{\gamma |p|}{m}))\,(\gamma^2 Y_3(\tfrac{\gamma |p|}{m})+v(0))^{1/2}, 
&n=0
\end{cases}
\nonumber
\end{align}
where $Y_\alpha(x)=\int_0^x |x-\sin\phi|^{-\alpha} d\phi$ and $n$ is the number of oscillations the fermions undergo before annihilation, see Fig. \ref{trajectory} in which the nodes of $z(t)$ correspond to $n=0,1,2$.

The time $t$ in Eq.(\ref{explicit-wf}) is the moment of the collision,
wheres $p=p_z$ is the electron momentum at the moment of creation. 
It should be chosen in such a way that
the condition $z(t)=0$ is satisfied, which makes $p$  a function of $t$ and $n$. 
Generally, $\Phi(0,t)$ is a sophisticated function of $t$,
which distinguishes the present problem from the positronium annihilation
where the fermion wavefunction exhibits simple oscillations $\propto \exp(-iE_{Ps}t)$, $E_{Ps}$ is the total positronium energy. 
However, the complicated $t$-dependence of the wave function reveals itself only at large intervals of time $\delta t\propto 1/\omega$, 
while the annihilation is a rapid process. Hence, the only contribution relevant to the annihilation process
originates from the part of wave function, which at the moment of collision oscillates with the frequency that matches the frequency $2\varepsilon$ of the photon pair in the final state. To extract this term we need to make the Fourier transform
\begin{align}
\Phi_{2\varepsilon}(0)=\int \exp(-2i\,\varepsilon t)\Phi(0,t)\,dt~.
\label{fourier}
\end{align}
The spin factor $\xi_\mathbf{p}$, which is present in (\ref{explicit-wf}), is omitted here because,
following conventional formalism, it is included in the definition of the cross section $\sigma(\varepsilon)$, which appears in (\ref{recoll}).

The integral in (\ref{fourier}) may be evaluated explicitly, which would 
in turn allow us to calculate the conversion coefficient $\Upsilon$ in (\ref{recoll}) numerically.
We elide these details here, instead using the wavefunction itself to
obtain a reliable analytic estimate for $\Upsilon$. 
A simple estimation for $\Upsilon$ may be obtained
via classical arguments. 
One takes into account that the $e^+e^-\rightarrow 2\gamma$ cross section
is of the order of $\sigma\sim \pi\,r_e^2$, see 
\cite{Page1957}, while
the typical separation of the wiggling fermion pair is $R\sim v/\omega\sim 1/\omega$, 
because the fermion velocity is $v\approx 1$. 
This leads one to the following estimate for the conversion coefficient
\begin{align}
\Upsilon_\text{clas}\,\sim\,\frac{\sigma}{R^2}\,\propto  \,\frac1{c^2}\,\omega^2 \sigma~.
\label{Ups-class}
\end{align}
However, this result grossly underestimates the probability of the recollision
because it does not take into account that the $e^-e^+$ pair propagates predominantly along the electric field.
We will see that this important property stems from the quantum 
nature of the problem.
To justify this claim we rewrite the conversion coefficient from Eq. (\ref{recoll}) 
as follows
\begin{align}
\Upsilon=4\langle v\sigma \rangle\int|\Phi_{2\varepsilon}(0)|^2\,\frac{d\varepsilon}{2\pi}.
\label{averaged s}
\end{align}
This equality can be considered as the definition of the averaged value 
$\langle v\sigma \rangle$ for the quantity $ |v|\sigma(\varepsilon)$. 
Using Parseval's theorem, we can rewrite (\ref{averaged s}) further
\begin{align}
\Upsilon=2\,\langle v\sigma \rangle \int| \Phi(0, t)|^2\,dt.
\label{averaged s0}
\end{align}
Change now the measure of integration  from $dt$ to $d|p|$. 
The necessary derivative $\dot{p}$ extracted from the known classical 
trajectory (\ref{trajectory_explicit}),
reads $|\dot{p}|=|v| F D^2(p)/(2n+v(0))$.
Calculating $| \Phi(0, t)|^2$ we presume that the annihilation processes, which take place at different $n$ do not interfere with each other since they are separated 
by large intervals of time.
To simplify the discussion, we also neglect the term with $n=0$ because careful analysis, which will be presented elsewhere,
shows that numerically its contribution is less than that of the $n=1$ term.
As a result, from (\ref{explicit-wf}), (\ref{averaged s0}) we derive
\begin{align}
\Upsilon_\text{quant} \,&\simeq \,\frac{\gamma\,F^{1/2}\,
\langle v\sigma \rangle }{2^{9/2}\,\ln^2(\pi/\gamma)}
\,\sum_{n\ge 1}\frac1{n^2}\,
\int_0^\infty\!\!
\exp\Big(\!-\frac {\pi\gamma^2 p^2}{2F}\Big)\,\frac{d|p|}{|v|}
\nonumber
\\
&\quad\quad\simeq
\frac{\pi^2}{192}\,\frac1{\hbar c^2}\,\frac{F}{\ln^2(\pi/\gamma)} \,\langle v\sigma \rangle\,.
\label{averaged s2}
\end{align}
The limits of integration over $|p|$ here take into account that the collision occurs only for negative values of $p$.
In the last step in (\ref{averaged s2}) we presumed that the collision velocity is large, $|v|\approx 1$, which is an adequate approximation. 

Equation  (\ref{averaged s2}) presents an appealing, important result.
The Planck constant
there  explicitly  shows its quantum origins, which contrasts
the classical nature of $\Upsilon_\text{clas}$ in (\ref{Ups-class}). 
In the static limit $\gamma=0$ the 
annihilation becomes impossible, $\Upsilon=0$, because the static electric field
can but only further separate the created fermions. Equation (\ref{averaged s2}) complies with
this condition, though for $\gamma\rightarrow 0$ 
the term  independent of the cross 
section decreases slowly, 
as $\propto \ln^{-2}(\pi/\gamma)$.

Most importantly, our quantum result (\ref{averaged s2}) greatly exceeds
the simple classical estimate (\ref{Ups-class})
\begin{align}
\frac{\Upsilon_\text{quant}}{\Upsilon_\text{clas}}\propto
\frac1{\gamma^2 \ln^2(\pi/\gamma)}
\frac{m^2}{F} \gg 1~,
\label{q/cl}
\end{align}
where we presumed that $\sigma \approx \langle v\sigma \rangle$. Observe a large factor $\gamma^{-2}$, which appears here 
(the proposed developments of laser facilities would make $\gamma$ small, $\gamma\sim 10^{-6}$).
The arguments presented above show that this impressive enhancement 
stems from the clear physical reason. 
In the coherent laser field the $e^-e^+$ pairs are produced with momenta, which 
are strongly aligned with the electric field. 

\begin{figure}[b]
{
\includegraphics[scale=0.6]{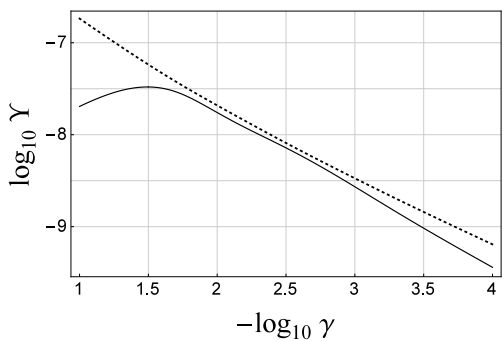}
}
\caption{
The number of photons per $e^-e^+$ pair $\Upsilon$ versus $\gamma=m\omega/F$ for $F=0.8\ m^2$;
dotted line - Eq. \eqref{averaged s2}, solid line - numerical results derived from
(\ref{recoll}), (\ref{explicit-wf}), (\ref{fourier}).}
\label{estimation}
\end{figure}

To study $\Upsilon_\text{quant}$ in more detail
we take the two-photon annihilation cross section for polarized fermions found in 
\cite{Page1957}, which for all energies can be approximated as $|v| \sigma(\varepsilon)\approx 
\pi r_e^2(m/\varepsilon)^2$ with accuracy $\ge 80\%$. 
The typical electron energy in the laser field is $\varepsilon \propto m/\gamma$. Therefore $\langle v\sigma \rangle\simeq \pi r_e^2\gamma^2$, which yields
\begin{align}
\Upsilon_\text{quant} \,\simeq \,
\frac{\pi^3}{192}\,\frac{r_e^2\,\gamma^2\,F}{\ln^2(\pi/\gamma)}~.
\label{Ups-quant}
\end{align}
This clear and attractive formula is one of the important 
results of this work.
We verified its validity by full scale numerical
calculations based on the formalism developed in 
Eqs.(\ref{recoll}), (\ref{explicit-wf}), and (\ref{fourier}). 
Our preliminary results are shown in Fig. \ref{estimation}.
Note that data derived from numerical analysis and from Eq.(\ref{Ups-quant})
are very close for a wide range of $\gamma$,
though the distinction grows for larger $\gamma$.  

Figure \ref{estimation} also shows that
at $F\approx m^2$ the maximal value of the conversion coefficient is $\Upsilon\approx 10^{-7}$.
This means that each $e^-e^+$ pair created in the laser field can produce $\sim 10^{-7}$
high energy photons. It should be emphasized that this is a very encouraging result,
as this probability of the photon production by several orders of magnitude exceeds the one predicted for the collision of a laser beam with the nucleus \cite{Kuchiev2007}. 
As was mentioned, the reason for the found enhancement stems from 
the coherent nature of the laser field.

\begin{figure}[h]
{\includegraphics[scale=0.3]{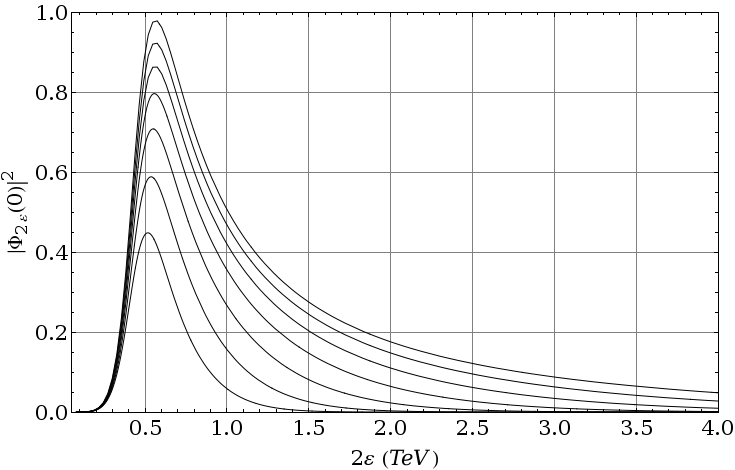}}
\caption{Fourier component of the wave function, $|\Phi_{2\varepsilon}(0)|^2$ (arbitrary units),  numerically calculated from Eq.(\ref{fourier}) for $n=0$ 
versus the electron energy $\varepsilon$ for $\gamma=2\cdot 10^{-6}$;
bottom to top: $F/m^2$ =  $ 0.0125, \ 0.025, \ 0.05, \ 0.1, \ 0.2, \ 0.4, \ 0.8$.}
 \label{energy}
\end{figure}

We can describe similarly the production of heavy, high energy particles, for example muons, triggered 
by the annihilation of $e^-e^+$ pairs created in the standing laser wave. 
To make this phenomenon possible $\gamma$ should be sufficiently small
to ensure that the wiggling energy of the $e^-$ and $e^+$ to exceed the mass $M$ of a heavy particle, 
$\varepsilon\sim m/\gamma>M$.
For $\gamma\sim 10^{-6}$ this wiggling energy is quite large, $\sim 1$ TeV, as 
one derives from Fig. \ref{energy}, where $|\Phi_{2\varepsilon}(0)|^2$ 
measures the probability that the $e^-e^+$ collision takes place with the energy $\varepsilon$.
The antenna mechanism considered gives a huge, exponential enhancement for such processes compared with the direct, 
Schwinger-type production of heavy particles. The probability for the latter is  $\propto \exp(-\pi M^2/F)$, 
whereas the antenna mechanism gives $\Upsilon \propto \exp(-\pi m^2/F)$, where $m$ is the electron mass. According to
Eq.(\ref{averaged s2}) the main suppression for $\Upsilon$ originates from the cross section, which 
decreases with energy. However, it is a power-type decrease, while the gain in probability is exponential, $\propto
\exp\big(\pi (M^2-m^2)/F\big)$.
Similarly, by comparing the production of high energy photons by an independent fermions $e^\pm$, 
which radiate due to their oscillation in the laser field,
and the photon production via $e^-e^+$ annihilation, considered in this work, we find that later process 
is exponentially enhanced compared to the former one.

Summarizing, it is demonstrated that in the foreseeable future 
high energy photons and massive particles with energies above $\sim 1$ TeV 
can be produced using the intense colliding laser beams.

This work was supported by the Australian Research Council.

\bibliographystyle{apsrev4-1}

\end{document}